\documentclass[journal,twoside,web]{ieeecolor}
\usepackage{tmi}
\usepackage{cite}
\usepackage{amsmath,amssymb,amsfonts}
\usepackage{algorithmic}
\usepackage{graphicx}
\usepackage{textcomp}
\usepackage{bm}
\usepackage[ruled]{algorithm2e}
\usepackage{booktabs}
\usepackage{multirow}

\def\BibTeX{{\rm B\kern-.05em{\sc i\kern-.025em b}\kern-.08em
    T\kern-.1667em\lower.7ex\hbox{E}\kern-.125emX}}
\begin{document}
\title{Diffusion Prior Regularized Iterative Reconstruction for Low-dose CT}
\author{Wenjun Xia, \IEEEmembership{Member, IEEE}, Yongyi Shi, Chuang Niu, Wenxiang Cong and Ge Wang, \IEEEmembership{Fellow, IEEE}
\thanks{This research is partially supported by National Cancer Institute (NCI) of the National Institutes of Health (NIH) grants R01CA237267, R01HL151561, R01EB031102 and R01EB032716.}
\thanks{W. Xia, Y. Shi, C. Niu, W. Cong and G. Wang are with the Department of Biomedical Engineering, School	of Engineering, Rensselaer Polytechnic Institute, Troy, 12180, NY, USA (e-mail: xiaw4@rpi.edu, shiy11@rpi.edu, niuc@rpi.edu, congw@rpi.edu, wangg6@rpi.edu). }
}

\maketitle

\begin{abstract}
Computed tomography (CT) involves a patient's exposure to ionizing radiation. To reduce the radiation dose, we can either lower the X-ray photon count or down-sample projection views. However, either of the ways often compromises image quality. To address this challenge, here we introduce an iterative reconstruction algorithm regularized by a diffusion prior. Drawing on the exceptional imaging prowess of the denoising diffusion probabilistic model (DDPM), we merge it with a reconstruction procedure that prioritizes data fidelity. This fusion capitalizes on the merits of both techniques, delivering exceptional reconstruction results in an unsupervised framework. To further enhance the efficiency of the reconstruction process, we incorporate the Nesterov momentum acceleration technique.  This enhancement facilitates superior diffusion sampling in fewer steps. As demonstrated in our experiments, our method offers a potential pathway to high-definition CT image reconstruction with minimized radiation.
\end{abstract}


\section{Introduction}
\label{sec:introduction}
Computed tomography (CT) has gained widespread utilities in diagnosis and therapy due to its high resolution, rapid scanning, and non-invasive nature. Despite the evolving landscape of medical imaging techniques, CT remains a pivotal modality. However, a prominent concern is the ionizing radiation associated with CT scanning. Such radiation effects can be accumulated in the body over time and potentially lead to radiation-induced diseases. Consequently, medical professionals adhere to the ``As Low As Reasonably 
Achievable'' (ALARA) principle when conducting CT scans, aiming to minimize the radiation exposure to patients. To enable CT radiation dose reduction, two prevalent strategies are (a) decreasing the number of X-ray photons in every projection and (b) down-sampling the number of projection views over a scan. Yet, both techniques can compromise the CT reconstruction quality, introducing severe noise and pronounced streaks in a reconstructed image. 
Given these issues, there are been on-going efforts made by researchers worldwide to enhance the image quality of low-dose CT scans. This is critical to ensure that patients optimally benefit from CT scanning.

At present, most commercial CT systems employ the filtered back-projection (FBP) algorithm for image reconstruction \cite{kak2001principles}. FBP intrinsically demands a sufficient amount of high-quality raw data for theoretically accurate results. When projection data from a low-dose CT scan are reconstructed using FBP, the resultant images suffer from both image noise and artifacts. 
To have the robustness of reconstructed images against noise and down-sampled data, iterative reconstruction (IR) algorithms were developed. These algorithms treat the CT measurement process as a linear system, and use iterative approximation techniques to solve the linear inverse problem. Notable among these are algebraic reconstruction technique (ART) \cite{gordon1970algebraic}, simultaneous iterative reconstructive technique (SIRT) \cite{trampert1990simultaneous}, and simultaneous algebraic reconstruction technique (SART) \cite{andersen1984simultaneous}. These methods tailored well-known optimization approaches to the CT problem. Impressively, the IR quality surpasses that of FBP in the cases of low-dose CT.

While IR algorithms represent a marked advancement over the FBP method, the quality of reconstructed images still struggles to meet clinical standards in many low-dose CT tasks. Consequently, building on the foundation of the IR algorithms, the model-based iterative reconstruction (MBIR) methods were developed \cite{thibault2007three, xia2023physics}. These algorithms construct a model that encompasses both fidelity and prior terms. The fidelity term draws on the statistical likelihood of the data, typically formulated within the data projection domain, where the theoretical model is well established. On the other hand, the prior term derives from empirical models rooted in human experience. This is typically set in the image domain to reflect our prior knowledge of underlying images to be reconstructed.
The prior terms are generally task-specific. For instance, Yu \textit{et al.} employed total variation (TV) regularization for CT reconstruction, yielding promising outcomes \cite{yu2005total}. Building on this, Niu \textit{et al.} expanded TV into a higher order version, introducing the total generalized variation (TGV), addressing the oversmoothing problem often associated with TV \cite{niu2014sparse}. In a bid to capitalize on the sparsity \cite{candes2006robust, donoho2006compressed} of CT images, Xu \textit{et al.} devised a learned redundant dictionary that produced excellent images \cite{xu2012low}. Recognizing the low-rank nature inherent in images, Gao \textit{et al.} employed nuclear norm regularization for superior image quality \cite{gao2011multi}. Furthermore, the realm of CT reconstruction has seen the advent of various other prior terms. These encompass techniques like nonlocal means filtering \cite{zhang2016spectral, chen2009bayesian}, tight wavelet frames \cite{gao2011multi}, transform learning \cite{chun2017sparse}, tensor dictionary learning \cite{wu2018low}, tensor factorization \cite{wu2020high}, and convolutional sparse coding \cite{bao2019convolutional}, among others. While the MBIR algorithm delivers further-improved imaging performance, secondary and residual artifacts could still be troublesome in sophisticated clinical applications. Also, the adaptability of MBIR is quite restricted, as the prior term and its weight must be crafted to suit specific scenarios. Finally, the cost of MBIR is high since many iterations are needed for the IR process to converge.

Over the past decade, the evolution of deep learning technology has been swift, finding applications across diverse sectors including image processing and computer vision \cite{lecun2015deep, goodfellow2016deep}. This growth has sparked innovation in medical imaging, leading to the emergence of various image reconstruction and analysis algorithms  \cite{wang2016perspective, wang2019machine}. Presently, deep learning-based CT imaging approaches can be roughly grouped into two categories: 1) the image-to-image method and 2) the data-to-image method \cite{xia2021magic}. 

The image-to-image method seeks to directly map corrupted CT images to their cleaner counterparts. An initial approach employed a three-layer convolutional neural network (CNN) targeting the removal of noise from low-dose CT images \cite{chen2017low}. Subsequent networks incorporated residual structures to increase the network depth, address the gradient vanishing issue,  and facilitate training convergence, which yielded encouraging outcomes \cite{chen2017lowdose, jin2017deep}. Kang \textit{et al.} \cite{kang2017deep} and Han \textit{et al.} \cite{han2016deep} further refined the U-Net architecture for low-dose CT imaging with enhanced performance. Yang \textit{et al.} integrated the generative adversarial network (GAN) \cite{goodfellow2020generative, gulrajani2017improved} and perceptual loss \cite{johnson2016perceptual} to align reconstruction closely with human visual perception \cite{yang2018low}. While these techniques are proficient in noise and artifact elimination, they cannot effectively recover information already lost in compromised images.

On the other hand, the data-to-image approach incorporates raw data directly in the network-based reconstruction in an end-to-end fashion so that the output aligns closely with the acquired data, thereby elevating the quality of the resultant images. Zhu \textit{et al.} introduced AUTOMAP to learn the transformation from raw data to images \cite{zhu2018image}. Recognizing the challenges in memory and computational efficiency of AUTOMAP, He \textit{et al.} introduced the trajectory-based transform \cite{he2020radon} and its down-sampled variant \cite{he2021downsampled}.
Another intriguing avenue involves leveraging a differentiable FBP to bridge the projection domain network with the image domain network, an approach shown to yield impressive results \cite{hu2020hybrid, zhang2021clear, tao2021learning}. In parallel, there has been growing interest in melding deep learning architectures with MBIR algorithms. Conceptually, this can be visualized as utilizing the neural network to offer prior information. Several methods embed the network within various MBIR algorithm frameworks, including the gradient descent algorithm \cite{chen2018learn, gupta2018cnn, xia2021magic}, primal-dual hybrid gradient (PDHG) algorithm \cite{adler2018learn}, alternating direction method of multipliers (ADMM) framework \cite{he2018optimizing}, and the fast iterative shrinkage-thresholding algorithm (FISTA) \cite{xiang2021fista}, etc.
These approaches have demonstrated significant potential. With the access to raw data, the outcomes are superior than that obtained only through image post-processing. However, a major critique remains that these methods work in a supervised learning mode. In clinical settings, obtaining paired data — necessary for supervised training — is practically rather difficult and expensive due to both additional workload and cost as well as patient safety and privacy. Furthermore, these techniques frequently produce results that are somewhat over-smoothed or hallucinated. Either losing true details or introducing false details can mislead radiologists in their interpretation of the resultant images \cite{WU2022100474, WU2022100475}.

Recently, the denoising diffusion probabilistic model (DDPM) has garnered significant attention \cite{ho2020denoising, song2020score}. Characterized by its generative process that transforms noise into a target image. The hallmark of DDPM is the flexibility and subtlety inherent in each Gaussian iteration. As a result, DDPM delivers exceptional visual performance, and enables diverse application including CT reconstruction \cite{xia2022low, xia2022patch}. Actually, both DDPM and MBIR operate iteratively from low-quality to high-quality images, bearing a remarkable similarity. This similarity suggests an opportunity to synergistically capitalize on the strengths of both methods.
Song \textit{et al.} \cite{song2022solve} leveraged interpolation in the projection domain during the DDPM inverse process, conditioning the sampled image on projection data. Chung \textit{et al.}, on the other hand, channeled sampled images to a manifold space prior for regularization, referred to as the manifold constrained gradient (MCG) method \cite{chung2022improve}. These innovative techniques achieve high-quality CT images via DDPM without the need for paired data, circumventing the clinical challenge of lacking such data. Indeed, the stellar performance of DDPM in image processing consistently yield results that visually surpass many contemporary alternatives. However, there is still room for improvement. For instance, Song's approach, which essentially conditions the sampled image through interpolation, might be susceptible to numerical inaccuracy and instability. In contrast, MCG, while conceptually resembling a likelihood constraint, adopts a gradient descent format in each iteration. This often means a protracted convergence, since the stochastic nature of diffusion could interfere the conditioning phase, and the final images may contain unintended random elements.

To synergize DDPM and MBIR, here we introduce the Diffusion Prior Regularized Iterative Reconstruction (DPR-IR) scheme. This innovative method seamlessly incorporates the iterative diffusion image as a priority in the iterative reconstruction framework. Eschewing the gradient descent used in MCG, our approach adopts the ordered-subset SART (OS-SART) \cite{wang2004ordered}, aiming for data fidelity and an expedited convergence rate.
Moreover, to shorten the DDPM iteration process, we integrate the Nesterov momentum acceleration technique \cite{beck2009fast}. In conjunction with the sampling methodology used in the denoising diffusion implicit model (DDIM) \cite{song2020denoising}, this Nesterov technique  significantly expedites the reconstruction process.
To secure numerical stability, we make improvements in the following two aspects.
First, rather than deriving the momentum from intermediate iterative results, we determine the momentum based on the estimate of the clean image at each iteration.
Second, we impose the TV constraint on the momentum to suppress
the sporadic jumps in the DDPM stochastic process.
Through these enhancements, our scheme sets a stage for more accurate, robust and efficient CT image reconstruction.
The pivotal contributions of our research include:
\begin{itemize}
	\item \textbf{Integration of DDPM and IR:} We have seamlessly embedded DDPM in the IR framework, optimizing its role in regularizing IR. This is implemented using a novel conditioning technique anchored on OS-SART, capitalizing IR results in an unsupervised deep learning mode.
	
	\item \textbf{Nesterov Momentum Acceleration:} Given the iterative nature of DDPM, we have devised an enhanced Nesterov momentum acceleration strategy, addressing specifically the intricacies of DDPM.
	
	\item \textbf{Empirical Validation:} Through rigorous experimentation, we have substantiated the efficacy of our proposed approach. This not only shows the potential of our approach but also opens a door to development of DDPM-based IR techniques for clinical applications.
\end{itemize}

\section{Methodology}
\label{sec:methodology}
\subsection{Diffusion Prior Regularized IR}
A universal model for the MBIR algorithm in CT reconstruction can be expressed in the following equation:
\begin{equation}
	\max_{\bm{x}} \log{p(\bm{x},\bm{y})}=\log{p(\bm{y}|\bm{x})} + \log{p(\bm{x})}.
	\label{eq:1}
\end{equation}
where $ \bm{y} $ denotes projection data, and $ \bm{x} $ represents an image to be derived, $ p(\bm{y}|\bm{x}) $ signifies the posterior probability of $ \bm{x} $ given the observed data $ \bm{y} $, and $ p(\bm{x}) $ describes the prior probability of $ \bm{x} $. Assuming that both $ \log{p(\bm{y}|\bm{x})} $ and $ \log{p(\bm{x})} $ are differentiable, the gradient descent algorithm can be employed to optimize Eq. (\ref{eq:1}). The iterative process to produce the final output can be outlined as
\begin{equation}
	\bm{x}_{k+1} = \bm{x}_{k} + \alpha_k \left[ \nabla_{\bm{x}} \log{p (\bm{y}|\bm{x}_k)} + \nabla_{\bm{x}} \log{p (\bm{x}_k)} \right],
	\label{eq:2}
\end{equation}
where $ \alpha_k $ corresponds to the step size in the $ k $-th iteration.

In the classic reconstruction paradigms, $ \log{p(\bm{y}|\bm{x})} $ is typically known and is differentiable. Consequently, the crux of the MBIR algorithm revolves around crafting $ \log{p(\bm{x})} $. It is worth noting that the explicit form of $ \log{p(\bm{x})} $ is often specialized and non-differentiable, demanding computational tricks. In this study, we will delve into the diffusion prior, showcasing its seamless integration with Eq. (\ref{eq:2}).

The DDPM consists of forward and reverse processes. The forward trajectory is defined by a Markov chain that progressively infuses Gaussian noise into each input CT image, articulated as
\begin{equation}
	q(\bm{x}_{1:T}|\bm{x}_{0}) = \prod_{t=1}^T q(\bm{x}_t|\bm{x}_{t-1}),
	\label{eq:3}
\end{equation}
where 
\begin{equation}
	q(\bm{x}_t|\bm{x}_{t-1}) = \mathcal{N} (\bm{x}_t| \sqrt{1 - \beta_t}\bm{x}_{t-1}, \beta_t \bm{I}),
	\label{eq:4}
\end{equation}
$ \bm{x}_{0} $ represents an uncorrupted CT image, and $ \beta_1, \beta_2, \cdots, \beta_T  $ constitute a sequence of predetermined variances. Leveraging the  properties of the Gaussian distribution, a direct sampling operation can be formulated at any chosen timestep $ t $:
\begin{equation}
	q(\bm{x}_t|\bm{x}_{0}) = \mathcal{N} (\bm{x}_t| \sqrt{\bar{\alpha}_t}\bm{x}_0, 	(1-\bar{\alpha}_t) \bm{I}),
	\label{eq:5}
\end{equation}
where $\alpha_t = 1-\beta_t$ and $ \bar{\alpha}_t = \prod_{i=1}^t \alpha_i $. As the iterative process goes, $ \bm{x}_T $ eventually becomes noise following a multi-dimensional Gaussian distribution. 

However, our objective is not to derive noise from the image but to reverse-engineer a realistic image under the data distribution of interest from a noise image under the Gaussian distribution. To this end, our focus is on the reverse diffusion process. This also follows a Markov chain trajectory aided by a neural network:
\begin{equation}
	p_{\theta} (\bm{x}_{0:T})=p(\bm{x}_{T})\prod_{t=1}^T 	p_{\theta}(\bm{x}_{t-1}|\bm{x}_{t}),
	\label{eq:6}
\end{equation}
where
\begin{equation}
	p_{\theta}(\bm{x}_{t-1}|\bm{x}_{t}) = 	\mathcal{N}(\bm{x}_{t-1}|\mu_{\theta}(\bm{x}_{t}, t), \sigma_t^2 \bm{I}),
	\label{eq:7}
\end{equation}
$ \bm{x}_{T} $ is drawn from the Gaussian distribution: $ p(\bm{x}_{T}) \sim \mathcal{N}(0, \bm{I}) $, the term $ \mu_{\theta} $ represents the expected value determined by a learned U-Net model as described in \cite{ho2020denoising}. Interestingly, in many applications, the preference leans towards deploying a U-Net to predict noise rather than a direct prediction of the mean. This choice is expressed as
\begin{equation}
	\mu_{\theta}(\bm{x}_{t}, t) = \frac{1}{\sqrt{\alpha_t}} \left(\bm{x}_{t}-\frac{\beta_t}{\sqrt{1-\bar{\alpha}_t}} \bm{\epsilon}_{\theta}(\bm{x}_{t}, t) \right),
	\label{eq:8}
\end{equation}
where $ \bm{\epsilon}_{\theta} $ signifies the noise prediction model, that is, the U-Net. To train this U-Net for noise prediction, we employ the following loss function:
\begin{equation}
	\mathcal{L} = \mathbb{E}_{\bm{x}_0}\mathbb{E}_{\bm{\epsilon}, t}  \left\| 	\bm{\epsilon} - \bm{\epsilon}_{\theta}(\sqrt{\bar{\alpha}_t}\bm{x}_{0}+\sqrt{1-\bar{\alpha}_t}\bm{\epsilon}, t) \right\|_2^2 .
	\label{eq:9}
\end{equation}
where $ \bm{\epsilon} $ denotes the noise and is sampled from a standard normal distribution: $ \bm{\epsilon} \sim \mathcal{N}(0, \bm{I}) $. To construct a desired image from the noise, the reverse process can be written as
\begin{equation}
	\bm{x}_{t-1} = \frac{1}{\sqrt{\alpha_t}} \left(\bm{x}_{t}-\frac{\beta_t}{\sqrt{1-\bar{\alpha}_t}} \bm{\epsilon}_{\theta}(\bm{x}_{t}, t) \right) + \sigma_t \bm{z}, 
	\label{eq:10}
\end{equation}
where $ \bm{z}\sim \mathcal{N}(0, \bm{I}) $ denotes a random variable sampled from the normal distribution.

\begin{figure}[t]
	\centering
	\includegraphics[width=1.0\linewidth]{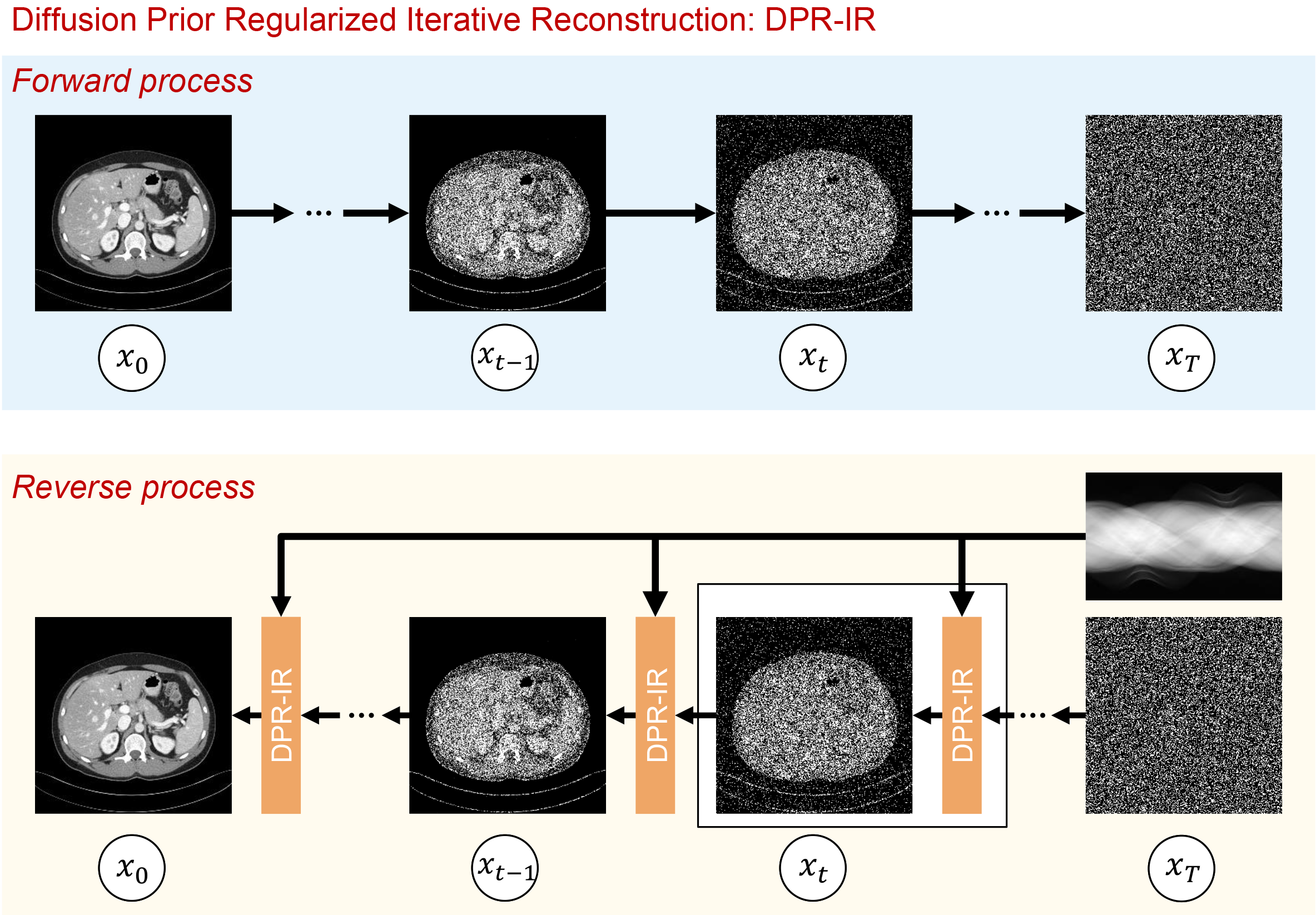}
	\caption{Illustration of the forward and reverse processes used for DPR-IR.}
	\label{fig:diffir1}
\end{figure}

\begin{algorithm}[t]
	\caption{Training the diffusion model $ \bm{\epsilon}_{\theta} $.}
	\label{alg:1}
	\KwIn{Variance schedule $ \beta_1, \beta_2, \cdots, \beta_T  $, high-quality image dataset $ p(\bm{x}) $}
	\KwOut{Trained model $ \bm{\epsilon}_{\theta} $}
	\BlankLine
	Initialize $ \bm{\epsilon}_{\theta} $ randomly
	
	\While{\textnormal{not converged}}{
		$ \bm{x}_0 \sim p(\bm{x}) $
		
		$ t\sim \mathrm{Uniform}(\{1,2,...,T\}) $
		
		$ \bm{\epsilon} \sim \mathcal{N}(0, \bm{I})$
		
		Update $\theta$ with the gradient $\nabla_{\theta} \left\| 	\bm{\epsilon} - \bm{\epsilon}_{\theta}(\sqrt{\bar{\alpha}_t}\bm{x}_{0}+\sqrt{1-\bar{\alpha}_t}\bm{\epsilon}, t) \right\|_2^2 $
	}
\end{algorithm}

Song \textit{et al.} demonstrated that the diffusion and reverse diffusion mechanisms of DDPM can be studied in a stochastic differential equation (SDE) framework. This advances the understanding of diffusion models, transitioning them from a discrete-time formulation to the continuous-time counterpart \cite{song2020score}. Specifically, the SDE representation for the diffusion process is given by
\begin{equation}
	\mathrm{d}\bm{x}=f(\bm{x},t)\mathrm{d}t+g(t)\mathrm{d}\bm{w},
	\label{eq:11}
\end{equation}
where $ \bm{w} $ represents the standard Wiener process, $ f(\cdot, t):\mathbb{R}^d\rightarrow \mathbb{R}^d $ serves as the drift function and $ g(\cdot): \mathbb{R}\rightarrow \mathbb{R}$ acts as the diffusion function. 
Song \textit{et al.} validated that by adopting the definitions for the drift and diffusion functions, the marginal distribution of the SDE aligns perfectly with the marginal distribution along the DDPM diffusion trajectory:
\begin{equation}
	\left\lbrace 
	\begin{aligned}
		&f(\bm{x},t) = -\frac{1}{2}\beta(t) \bm{x},\\
		& g(t) = \sqrt{\beta(t)}.
	\end{aligned}
	\right.	
	\label{eq:12}
\end{equation}
Indeed, by specifying the drift and diffusion functions properly, a variety of diffusion processes can be obtained, all of which share the same essential characteristics yet with different computational details. Song \textit{et al.} further presented the SDE for the reverse diffusion process:
\begin{equation}
	\mathrm{d}\bm{x}=\left[f(\bm{x},t) -g(t)^2 \nabla_{\bm{x}} \log{p (\bm{x})} \right]\mathrm{d}t+g(t)\mathrm{d}\bar{\bm{w}},
	\label{eq:13}
\end{equation}
where $ \bar{\bm{w}} $ is the Wiener process for the reverse SDE, and the term $ \nabla_{\bm{x}} \log{p (\bm{x})} $ is the score function, aligning closely with the gradient of the prior term. The noise prediction model, $ \bm{\epsilon}_{\theta} (\bm{x},t) $, is congruent to a scaled score function; i.e., $ \bm{\epsilon}_{\theta} (\bm{x}_t,t)=-\sigma_t \nabla_{\bm{x}} \log{p (\bm{x}_t)} $, as illustrated in \cite{lu2022dpm}. Based on the above analyses, DDPM learns the inherent prior distribution of data rigorously. This insight underscores the critical role of these diffusion models in modeling prior knowledge in a data-driven fashion and enabling the maximum a posteriori (MAP) reconstruction, a holy grail of the tomographic reconstruction field. 

After a data distribution is captured by DDPM as the most comprehensive prior information, it can be incorporated into the IR framework for ``plug and play'' \cite{venkatakrishnan2013plug}. Along this direction, Song's approach first undertakes conditioning in the projection domain and subsequently leverages the FBP for transition into the image domain \cite{song2022solve}, and Chung's method employs the gradient descent framework for conditioning \cite{chung2022improve}. However, these methods seem suboptimal in enforcing data constraints and ensuring fast convergence. 
To address these challenges, in our study we use the popular OS-SART algorithm \cite{wang2004ordered} for optimizing the fidelity term. OS-SART targets data consistency and boasts rapid convergence, directly addressing the limitations of the aforementioned methods. Specifically, during the reverse diffusion process at timestep $ t $, $ \bm{x}_t $ serves as the initialization for the OS-SART that then produces a data-consistent result. The forward and reverse mechanisms can be seen in Fig. \ref{fig:diffir1}. The training procedure for our new model is summarized in Algorithm \ref{alg:1}.  The iterative reconstruction process that takes advantage of DDPM is outlined in Algorithm \ref{alg:2}, which is named as DPR-IR-I.

\begin{algorithm}[t]
	\caption{Inferencing with DPR-IR-I.}
	\label{alg:2}
	\KwIn{Variance schedule $ \beta_1, \beta_2, \cdots, \beta_T  $, trained model $ \bm{\epsilon}_{\theta} $, corrupted projection dataset $ p(\bm{y}) $}
	\KwOut{$ \bm{x}_0 $}
	\BlankLine
	$ \bm{y} \sim p(\bm{y}) $
	
	$ \bm{x}_T \sim \mathcal{N}(0, \bm{I}) $
	
	\For{$t=T,...,1$}{
		$ \tilde{\bm{x}}_t = $ OS-SART$(\bm{x}_t, \bm{y}) $
		
		$\bm{z_t} \sim \mathcal{N}(0, \bm{I})$ if $t>1$ else $\bm{z_t}=0$
		
		$\bm{x}_{t-1} = \frac{1}{\sqrt{\alpha_t}} \left(\tilde{\bm{x}}_t-\frac{\beta_t}{\sqrt{1-\bar{\alpha}_t}} \bm{\epsilon}_{\theta}(\tilde{\bm{x}}_t, t) \right) + \sigma_t \bm{z_t}$
		
	}
\end{algorithm}

\subsection{Nesterov Momentum Acceleration}
DDPM offers a prior for IR, and is amendable to the Nesterov momentum acceleration which is commonly utilized in IR algorithms. This inclusion addresses the long computational time for DDPM iterations. In traditional Nesterov momentum acceleration applications, the momentum is applied to the intermediate iteration result $ \bm{x}_t $. For DDPM, however, applying this to intermediate iteration results can cause temporal mismatches. This temporal misalignment between images at various timesteps has an adverse effect on achieving optimal outcomes.
In response to this challenge, an alternative strategy we propose involves estimating the clean image during each iteration. Then, 
the momentum is applied directly to the estimated clean image. This approach is feasible because Eq. (\ref{eq:10}) can be deconstructed into the two distinct phases:
\begin{equation}
	\left\lbrace 
	\begin{aligned}
		&\hat{\bm{x}}_{0|t} = \frac{1}{\sqrt{\bar{\alpha}_t}} \left( \bm{x}_t - \sqrt{1-\bar{\alpha}_t}\bm{\epsilon}_{\theta}(\bm{x}_{t}, t)\right),\\
		& \bm{x}_{t-1} = \frac{\sqrt{\bar{\alpha}_{t-1}}\beta_t}{1-\bar{\alpha}_t} \hat{\bm{x}}_{0|t} + \frac{\sqrt{\alpha_{t}}(1-\bar{\alpha}_{t-1})}{1-\bar{\alpha}_t} \bm{x}_{t} + \sigma_t \bm{z}_t,
	\end{aligned}
	\right.	
	\label{eq:14}
\end{equation}
where $ \hat{\bm{x}}_{0|t} $ represents an estimate of $ \bm{x}_0 $ derived from an intermediate result at timestep $t$. By adopting this strategy, we can seamlessly apply momentum acceleration to $ \hat{\bm{x}}_{0|t} $ while circumventing the mismatch between the network input and the corresponding timestep.

Our objective in integrating the momentum is to cut down the number of iterations, thus improving the efficiency of the reconstruction process. To do so, we introduce the denoising diffusion implicit model (DDIM) \cite{song2020denoising}. On its non-Markovian basis, DDIM allows the use of more substantial step sizes. Furthermore, as the forward diffusion process of DDIM is the same as that of DDPM, there is no need for network retraining. The sampling algorithm for DDIM is characterized as follows:
\begin{equation}
	\left\lbrace 
	\begin{aligned}
		&\hat{\bm{x}}_{0|t} = \frac{1}{\sqrt{\bar{\alpha}_t}} \left( \bm{x}_t - \sqrt{1-\bar{\alpha}_t}\bm{\epsilon}_{\theta}(\bm{x}_{t}, t)\right),\\
		& \bm{x}_{t-1} = \sqrt{\bar{\alpha}_{t-1}} \hat{\bm{x}}_{0|t} +
		\sqrt{1-\bar{\alpha}_{t-1}-\sigma_t^2} \bm{\epsilon}_{\theta}(\bm{x}_{t}, t) \bm{x}_{t} + \sigma_t \bm{z}_t,
	\end{aligned}
	\right.	
	\label{eq:15}
\end{equation}
Thus, the DDIM-integrated DPR-IR without Nesterov momentum acceleration is referred to as DPR-IR-II. And the variants of DPR-IR-II employing Nesterov momentum acceleration can be encapsulated in Algorithms \ref{alg:3} and \ref{alg:4}  
with the momentum acceleration based on $ \bm{x}_t $ (used in the conventional momentum) and $ \hat{\bm{x}}_{0|t} $ (estimated for computing our modified momentum) respectively. These two algorithms are denoted as DPR-IR-III and DPR-IR-IV respectively.

\begin{algorithm}[t]
	\caption{DPR-IR-III.}
	\label{alg:3}
	\KwIn{Variance schedule $ \beta_1, \beta_2, \cdots, \beta_T  $, trained model $ \bm{\epsilon}_{\theta} $, corrupted projection dataset $ p(\bm{y}) $}
	\KwOut{$ \bm{x}_0 $}
	\BlankLine
	
	$ \tau \subset [1,2,...,T] $ with a length of $ S<T $
	
	$ \bm{y} \sim p(\bm{y}) $
	
	$ \bm{r}_{\tau_S} =\bm{x}_{\tau_S} \sim \mathcal{N}(0, \bm{I}) $
	
	$ \eta=1 $
	
	\For{$j=S,...,1$}{
		$ \tilde{\bm{x}}_{\tau_j} = $ OS-SART$(\bm{r}_{\tau_j}, \bm{y}) $
		
		$ \hat{\bm{x}}_{0|\tau_{j}} = \frac{1}{\sqrt{\bar{\alpha}_{\tau_j}}} \left( \tilde{\bm{x}}_{\tau_j} - \sqrt{1-\bar{\alpha}_{\tau_j}}\bm{\epsilon}_{\theta}(\tilde{\bm{x}}_{\tau_j}, {\tau_j})\right) $
		
		$ \bar{\bm{x}}_{\tau_{j-1}} = \sqrt{\bar{\alpha}_{\tau_{j-1}}} \hat{\bm{x}}_{0|\tau_{j}} +
		\sqrt{1-\bar{\alpha}_{\tau_{j-1}}-\sigma_{\tau_j}^2} \bm{\epsilon}_{\theta}(\tilde{\bm{x}}_{\tau_j}, \tau_j) \tilde{\bm{x}}_{\tau_j}$
		
		$ \eta' :\leftarrow (1+\sqrt{1+4\eta^2})/2 $
		
		$\bm{r}_{\tau_{j-1}} = \bar{\bm{x}}_{\tau_{j-1}} + \frac{\eta - 1}{\eta'}\left(\bar{\bm{x}}_{\tau_{j-1}} - \bar{\bm{x}}_{\tau_{j}}\right)$
		
		$ \eta:\leftarrow  \eta'$
		
		$\bm{z}_{\tau_j} \sim \mathcal{N}(0, \bm{I})$ if $t>1$ else $\bm{z}_{\tau_j}=0$
		
		$ \bm{x}_{\tau_{j-1}} = \bar{\bm{x}}_{\tau_{j-1}} +\sigma_{\tau_j} \bm{z}_{\tau_j} $
		
	}
\end{algorithm}

\begin{algorithm}[t]
	\caption{DPR-IR-IV.}
	\label{alg:4}
	\KwIn{Variance schedule $ \beta_1, \beta_2, \cdots, \beta_T  $, trained model $ \bm{\epsilon}_{\theta} $, corrupted projection dataset $ p(\bm{y}) $}
	\KwOut{$ \bm{x}_0 $}
	\BlankLine
	
	$ \tau \subset [1,2,...,T] $ with a length of $ S<T $
	
	$ \bm{y} \sim p(\bm{y}) $
	
	$ \bm{x}_{\tau_S} \sim \mathcal{N}(0, \bm{I}) $
	
	$ \eta=1 $
	
	\For{$j=S,...,1$}{
		$ \tilde{\bm{x}}_{\tau_j} = $ OS-SART$(\bm{x}_{\tau_{j}}, \bm{y}) $
		
		$ \hat{\bm{x}}_{0|\tau_{j}} = \frac{1}{\sqrt{\bar{\alpha}_{\tau_j}}} \left( \tilde{\bm{x}}_{\tau_j} - \sqrt{1-\bar{\alpha}_{\tau_j}}\bm{\epsilon}_{\theta}(\tilde{\bm{x}}_{\tau_j}, {\tau_j})\right) $
		
		$ \eta' :\leftarrow (1+\sqrt{1+4\eta^2})/2 $
		
		$\bm{r}_{\tau_{j-1}} = \hat{\bm{x}}_{0|\tau_{j}} + \frac{\eta - 1}{\eta'}\left(\hat{\bm{x}}_{0|\tau_{j}} - \hat{\bm{x}}_{0|\tau_{j+1}}\right)$
		
		$ \eta:\leftarrow  \eta'$
		
		$ \bar{\bm{x}}_{\tau_{j-1}} = \sqrt{\bar{\alpha}_{\tau_{j-1}}} \bm{r}_{\tau_{j-1}} +
		\sqrt{1-\bar{\alpha}_{\tau_{j-1}}-\sigma_{\tau_j}^2} \bm{\epsilon}_{\theta}(\tilde{\bm{x}}_{\tau_j}, \tau_j) \tilde{\bm{x}}_{\tau_j}$

		$\bm{z}_{\tau_j} \sim \mathcal{N}(0, \bm{I})$ if $t>1$ else $\bm{z}_{\tau_j}=0$
		
		$ \bm{x}_{\tau_{j-1}} = \bar{\bm{x}}_{\tau_{j-1}} +\sigma_{\tau_j} \bm{z}_{\tau_j} $
		
	}
\end{algorithm}

In the DDPM implementations, two successive iterations might contain random changes, leading to a non-smooth momentum estimate and distorting the final image. To address this, building upon DPR-IR-IV we further propose constraining the momentum using total variation (TV). By leveraging TV, the momentum can enhance image quality in a more refined manner, averting potential degradation by erratic momentum shifts. In this study, we utilize the Chambolle-Pock projection algorithm \cite{chambolle2011first} for this purpose. The whole workflow is encapsulated in Algorithm \ref{alg:5}.

%\begin{figure*}[htbp]
%	\centering
%	\includegraphics[width=0.6\linewidth]{figures/DiffIR2.png}
%	\caption{Visual representation of the forward and reverse mechanisms in DPR-IR.}
%	\label{fig:diffir2}
%\end{figure*}

\begin{algorithm}[t]
	\caption{DPR-IR-V.}
	\label{alg:5}
	\KwIn{Variance schedule $ \beta_1, \beta_2, \cdots, \beta_T  $, trained model $ \bm{\epsilon}_{\theta} $, corrupted projection dataset $ p(\bm{y}) $}
	\KwOut{$ \bm{x}_0 $}
	\BlankLine
	
	$ \tau \subset [1,2,...,T] $ with a length of $ S<T $
	
	$ \bm{y} \sim p(\bm{y}) $
	
	$ \bm{x}_{\tau_S} \sim \mathcal{N}(0, \bm{I}) $
	
	$ \eta=1 $
	
	\For{$j=S,...,1$}{
		$ \tilde{\bm{x}}_{\tau_j} = $ OS-SART$(\bm{x}_{\tau_{j}}, \bm{y}) $
		
		$ \hat{\bm{x}}_{0|\tau_{j}} = \frac{1}{\sqrt{\bar{\alpha}_{\tau_j}}} \left( \tilde{\bm{x}}_{\tau_j} - \sqrt{1-\bar{\alpha}_{\tau_j}}\bm{\epsilon}_{\theta}(\tilde{\bm{x}}_{\tau_j}, {\tau_j})\right) $
		
		$ \bm{m}_{\tau_{j}} =$Chambolle-Pock$ \left(\hat{\bm{x}}_{0|\tau_{j}} - \hat{\bm{x}}_{0|\tau_{j+1}}\right) $
		
		$ \eta' :\leftarrow (1+\sqrt{1+4\eta^2})/2 $
		
		$\bm{r}_{\tau_{j-1}} = \hat{\bm{x}}_{0|\tau_{j}} + \frac{\eta - 1}{\eta'}\bm{m}_{\tau_{j}}$
		
		$ \eta:\leftarrow  \eta'$
		
		$ \bar{\bm{x}}_{\tau_{j-1}} = \sqrt{\bar{\alpha}_{\tau_{j-1}}} \bm{r}_{\tau_{j-1}} +
		\sqrt{1-\bar{\alpha}_{\tau_{j-1}}-\sigma_{\tau_j}^2} \bm{\epsilon}_{\theta}(\tilde{\bm{x}}_{\tau_j}, \tau_j) \tilde{\bm{x}}_{\tau_j}$

		$\bm{z}_{\tau_j} \sim \mathcal{N}(0, \bm{I})$ if $t>1$ else $\bm{z}_{\tau_j}=0$
		
		$ \bm{x}_{\tau_{j-1}} = \bar{\bm{x}}_{\tau_{j-1}} +\sigma_{\tau_j} \bm{z}_{\tau_j} $
		
	}
\end{algorithm}

\begin{table*}[htbp]
	\caption{Quantitative Results (MEAN$\pm$SD) on the Testing Data.}
	\label{tab:1}
	\centering
	\begin{tabular}{lcccccc}
		\toprule
		\multirow{2}{*}{Method}  & \multicolumn{3}{c}{\%10 Dose Data} & \multicolumn{3}{c}{96-View Data}  \\
		\cmidrule{2-4}
		\cmidrule{5-7}
		& PSNR                 & SSIM                 & RMSE                 & PSNR                 & SSIM                 & RMSE                 \\
		\midrule
		
		FBP             &          32.98$ \pm $1.48          &       0.7006$ \pm $0.0641               &          0.0228$ \pm $0.0037            &         24.46$ \pm $0.52             &          0.3466$ \pm $0.0259            &          0.0600$ \pm $0.0036            \\
		MCG (GD)  &          23.86$ \pm $0.66            &           0.7387$ \pm $0.0229          &         0.0643$ \pm $0.0049             &          26.03$ \pm $0.67            &          0.7675$ \pm $0.0212            &          0.0501$ \pm $0.0039           \\

		DPR-IR-I  &          40.19$ \pm $0.61            &           0.9332$ \pm $0.0109           &         0.0098$ \pm $0.0007             &          37.28$ \pm $0.72            &          0.9065$ \pm $0.0107            &          0.0137$ \pm $0.0011            \\

        DPR-IR-II &          40.30$ \pm $0.55            &           0.9377$ \pm $0.0080          &         0.0097$ \pm $0.0006             &          38.36$ \pm $0.78            &          0.9274$ \pm $0.0078            &          0.0121$ \pm $0.0011           \\
		
		DPR-IR-III  &          40.67$ \pm $0.60           &           0.9380$ \pm $0.0082          &         0.0093$ \pm $0.0006             &          39.90$ \pm $0.73            &          0.9333$ \pm $0.0082            &          0.0102$ \pm $0.0008            \\
		DPR-IR-IV  &          39.41$ \pm $0.56          &           0.9263$ \pm $0.0097          &         0.0107$ \pm $0.0007             &          36.96$ \pm $0.98            &          0.8976$ \pm $0.0192            &          0.0143$ \pm $0.0016            \\
		DPR-IR-V &          40.04$ \pm $0.53            &        0.9328$ \pm $0.0088              &        0.0100$ \pm $0.0006              &         38.55$ \pm $0.65             &         0.9236$ \pm $0.0085             &        0.0119$ \pm $0.0008   \\   
		\bottomrule       
	\end{tabular}
\end{table*}

\begin{figure*}[t]
	\centering
	\includegraphics[width=0.9\linewidth]{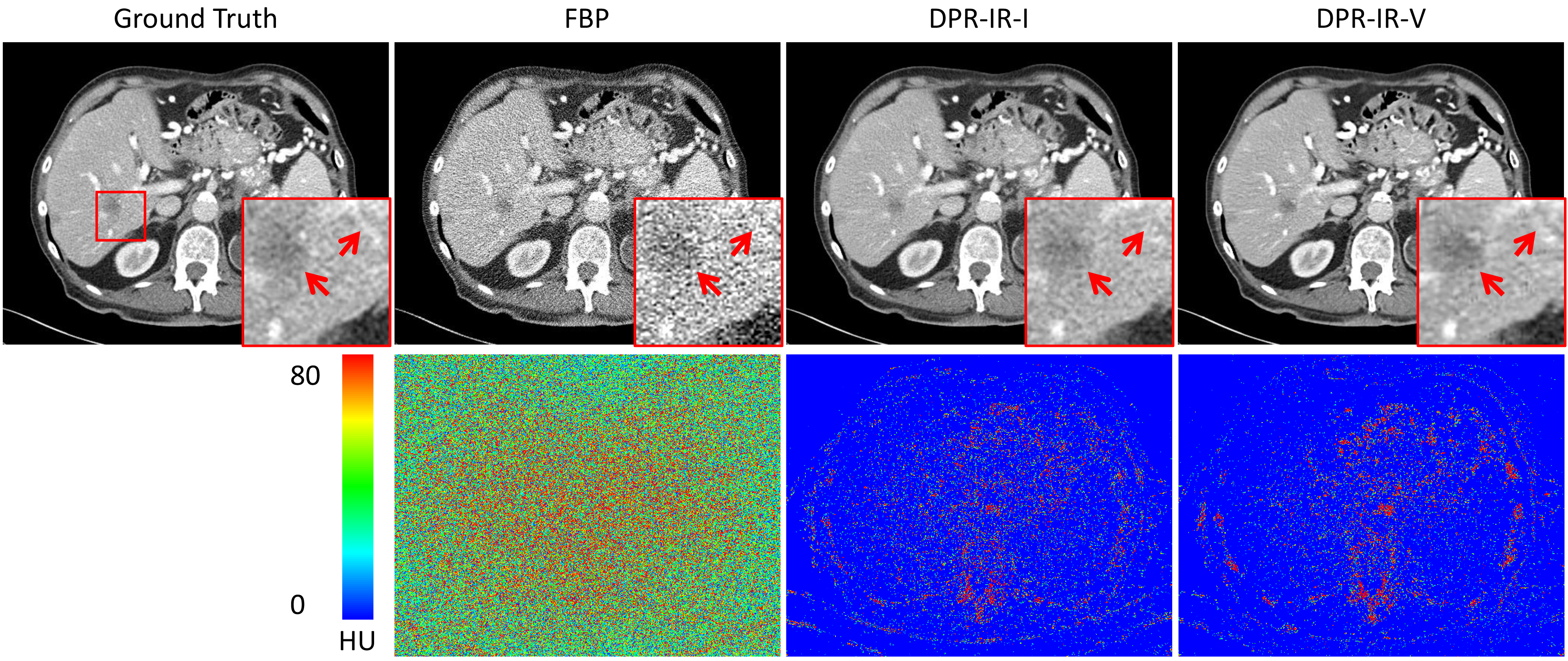}
	\caption{Abdominal reconstructions and their associated absolute difference maps using DPR-IR-I and V with 10\% dose data. The display window is [-160, 240] HU.}
	\label{fig:res_lowdose}
\end{figure*}

\begin{figure*}[t]
	\centering
	\includegraphics[width=0.9\linewidth]{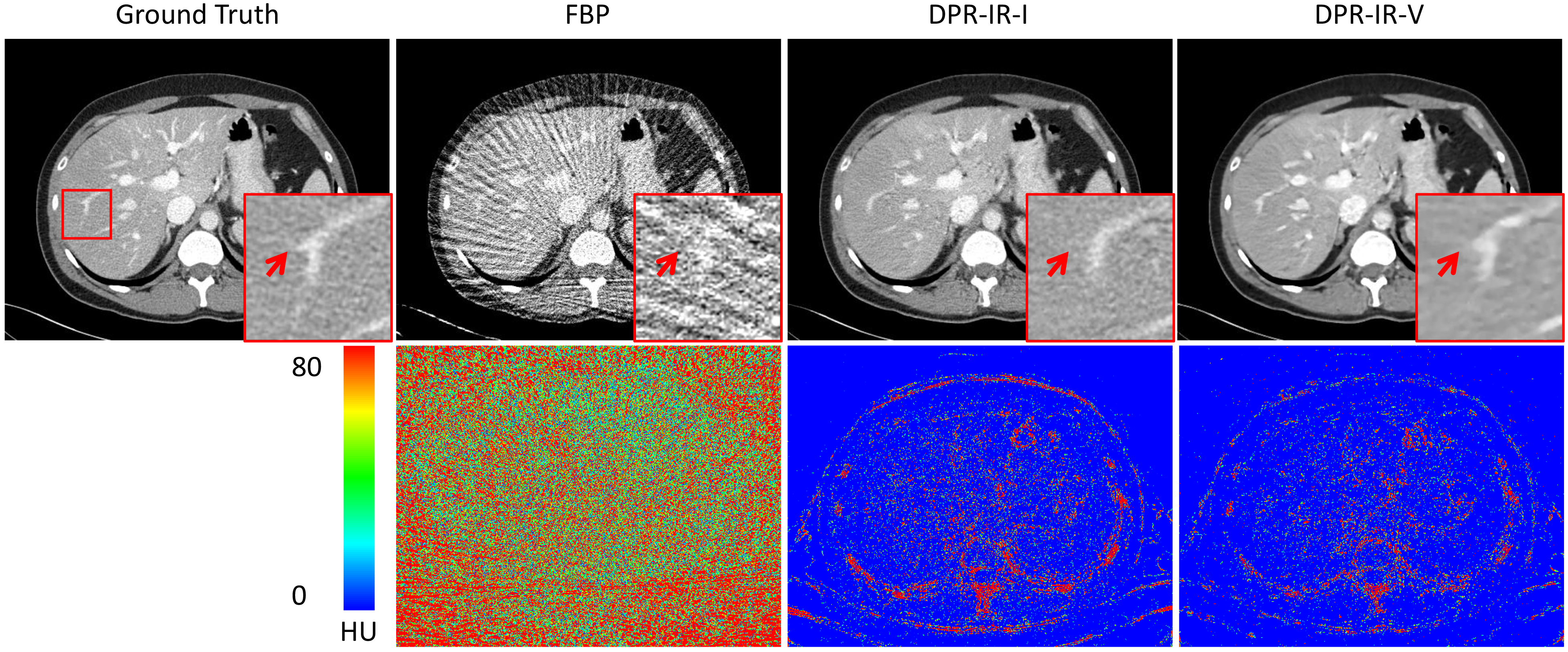}
	\caption{Abdominal reconstructions and their associated absolute difference maps using DPR-IR-I and V with 96-view data. The display window is [-160, 240] HU.}
	\label{fig:res_sparseview}
\end{figure*}

\section{Experiments And Results}
\label{sec:experiments}
\subsection{Experimental Setup}
To validate the efficacy of DPR-IR-V, we employ DPR-IR-I — IV to highlight the significance of each innovative step.

In our experiments, we sourced the dataset from "the 2016 NIH-AAPM-Mayo Clinic Low-Dose CT Grand Challenge". The AAPM dataset encompasses 2,378 full-dose CT images with a thickness of 3mm from 10 distinct patients. From this pool, 1,923 images obtained from 8 patients were utilized as the training set. The remaining 455 images from the other 2 patients were reserved to test the imaging performance.

With these image data, we performed image reconstruction using competing reconstruction algorithms. The imaging geometry was defined with the source-to-center distance of 595 mm, and the source-to-detector distance of 1085.6 mm, 
 image size of 512$ \times $512, pixel size of 0.6641 mm, the arc detector array of 736 elements, and detector element of 1.2858 mm. To further enhance the realism of the simulation, we introduced additional noise to the projections as follows \cite{niu2014sparse}:
\begin{equation}
	I = \mathcal{P}(I_0 \exp{-\bm{y}}) + \mathcal{N}(0, \sigma_e^2),
\end{equation}
where $ \mathcal{P}(\cdot) $ represents the Poisson distribution, $ I_0 $ denotes the count of incident photons, and $ \sigma_e^2 $ stands for the variance of electronic noise. In the context of this study, the electronic noise level was pre-defined as $ \sigma_e^2=10 $. Additionally, an incident photon count of $ I_0=1\mathrm{e}6 $ was used as the standard dose. Our simulation was conducted under the two conditions. First, we established $ I_0 $ at \%10 of the standard dose. This enabled us to have low-dose data from 1,024 different views over a full scan range. Second, we elevated $ I_0 $ to standard dose to acquire sparse view data from 96 angles uniformly sampling a 360-degree range.

\begin{figure*}[htbp]
	\centering
	\includegraphics[width=0.9\linewidth]{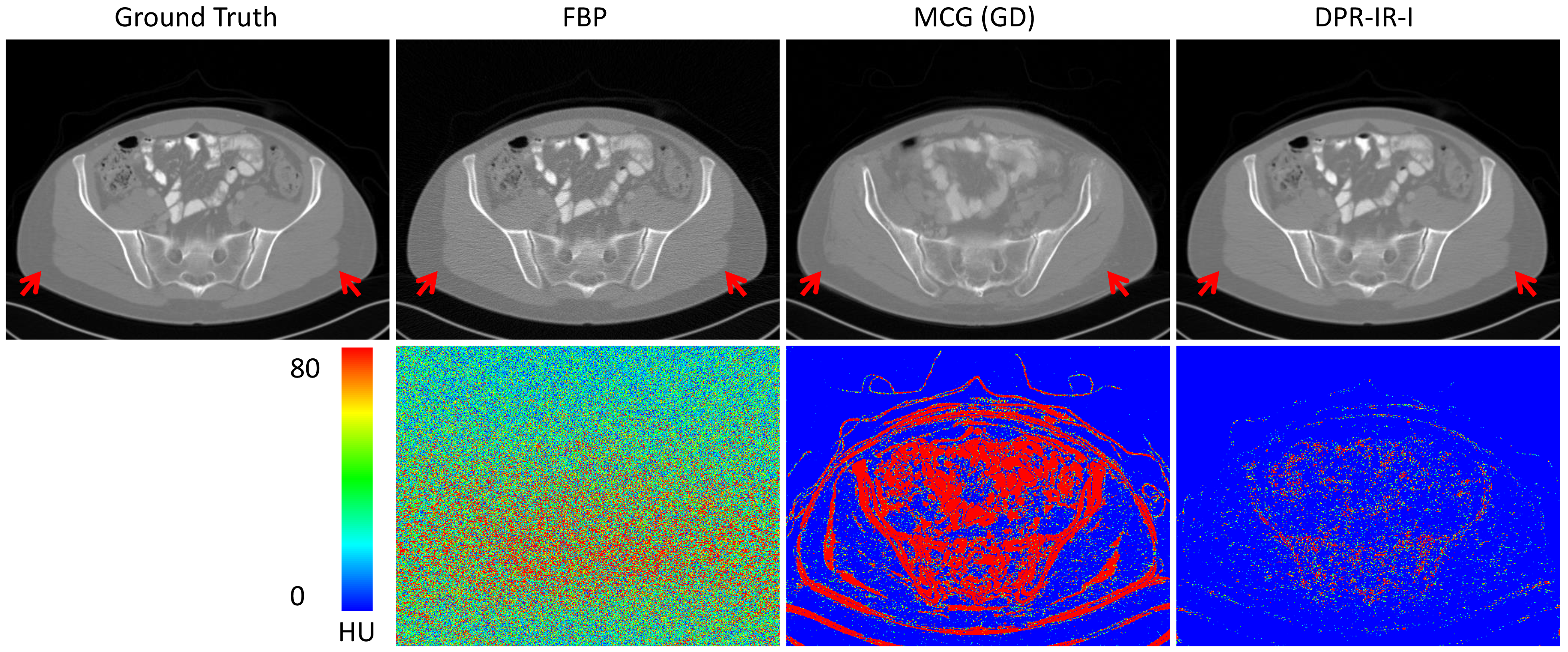}
	\caption{Pelvis reconstructions and their associated absolute difference maps using different conditioning methods and 10\% dose data. The display window is [-1000, 1000] HU.}
	\label{fig:res_sart_lowdose}
\end{figure*}

\begin{figure*}[t]
	\centering
	\includegraphics[width=0.9\linewidth]{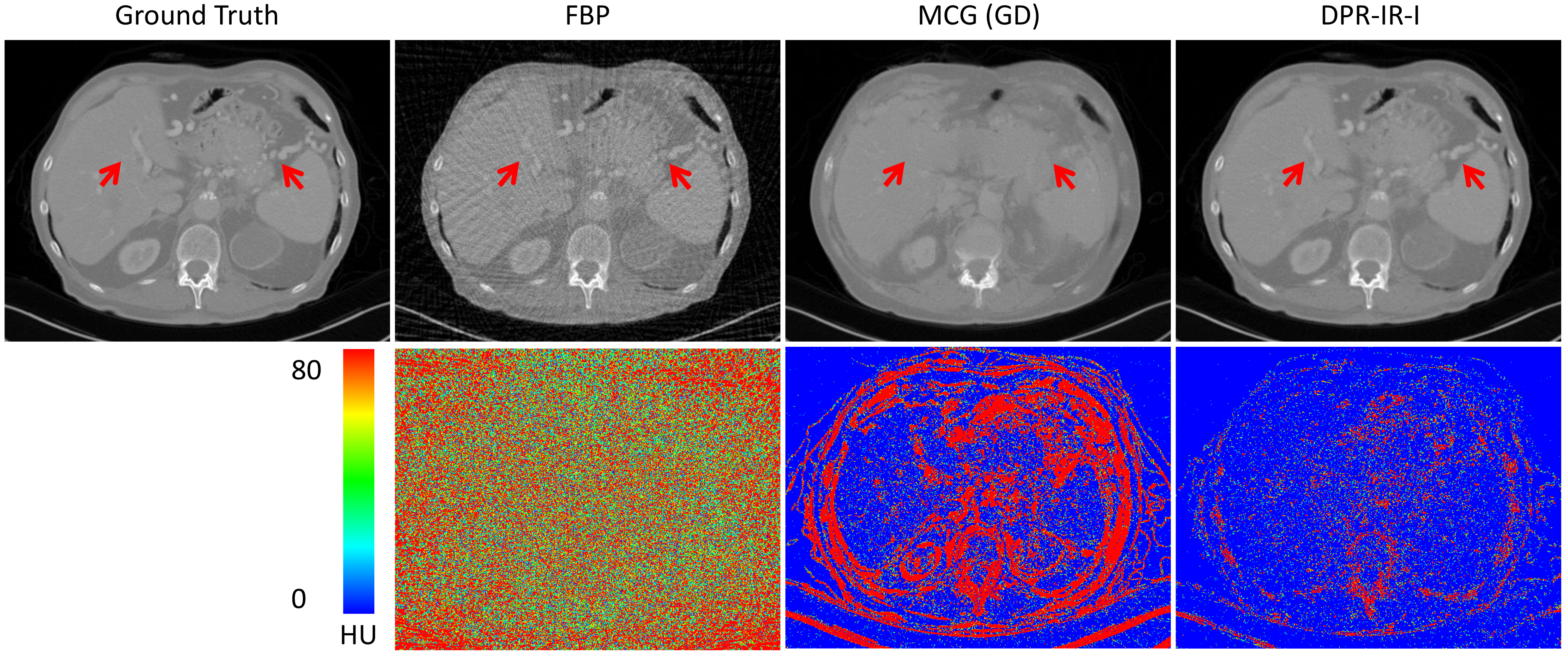}
	\caption{Abdominal reconstructions and their associated absolute difference maps using different conditioning methods and 96-view data. The display window is [-1000, 1000] HU.}
	\label{fig:res_sart_sparseview}
\end{figure*}

In the execution of DDPM, we employed a linear sequence for the variance schedule, with the starting and ending values of $ \beta_1 = 1\mathrm{e}-4 $ and $ \beta_T = 0.02 $ respectively. While DDPM was set to undergo 1,000 timesteps for DPR-IR-I, for the accelerated variant the DDIM was configured to run in only 200 steps for DPR-IR-II, III, IV and V respectively.

\subsection{Quantitative Metrics}

Table \ref{tab:1} displays the quantitative  evaluation results on different algorithms, specifically detailing the mean values and standard deviations (SDs) for the common metrics PSNR, SSIM, and RMSE. The MCG (GD) method \cite{chung2022improve} underwent a transformation from its score SDE iteration to the DDPM format. This method utilizes the gradient descent (GD) technique for conditioning. In terms of the aforementioned classic metrics, DPR-IR-III delivered the more favorable results on both the 10\% dose data and the 96-view data but these metrics are often not consistent to radiologists' visual impression. The following subsections will provide more analyses of these visual results.

\subsection{Acceleration Results}
Figs. \ref{fig:res_lowdose} and \ref{fig:res_sparseview} display abdominal reconstructions and their corresponding absolute difference maps using DPR-IR-I and V for 10\% dose data and 96-view data, respectively.

In Fig. \ref{fig:res_lowdose}, both DPR-IR-I and V effectively mitigate noise in the FBP reconstruction. Notably, their visual outcomes closely resemble the ground truth. The DPR-IR-V performance aligns well with that of DPR-IR-I. Within the magnified region of interest (ROI), the DPR-IR-V result unveils fine structures more clearly, especially in the regions pinpointed by the red arrows, than that with DPR-IR-I. The difference maps illustrate that DPR-IR-V gave marginally more noise than DPR-IR-I but within an acceptable range.

\begin{figure*}[htbp]
	\centering
	\includegraphics[width=1.0\linewidth]{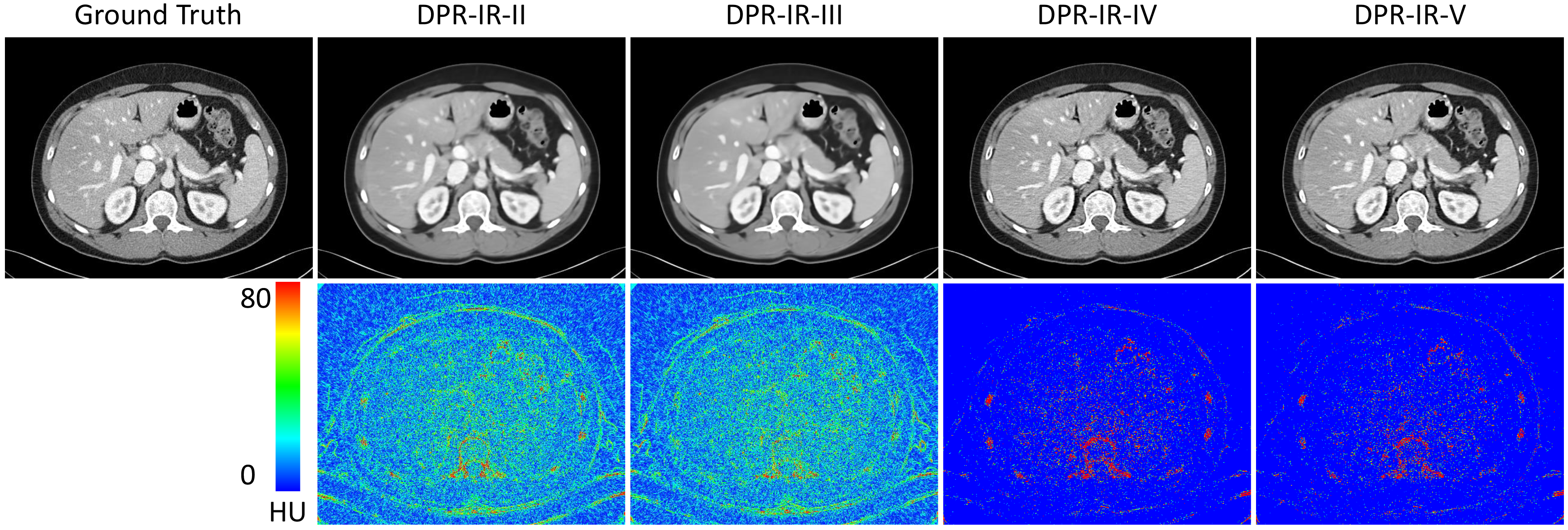}
	\caption{Abdominal reconstructions and their associated absolute difference maps using 200 sampling steps with 10\% dose data. The display window is [-160, 240] HU.}
	\label{fig:res_mom_lowdose}
\end{figure*}

\begin{figure*}[t]
	\centering
	\includegraphics[width=1.0\linewidth]{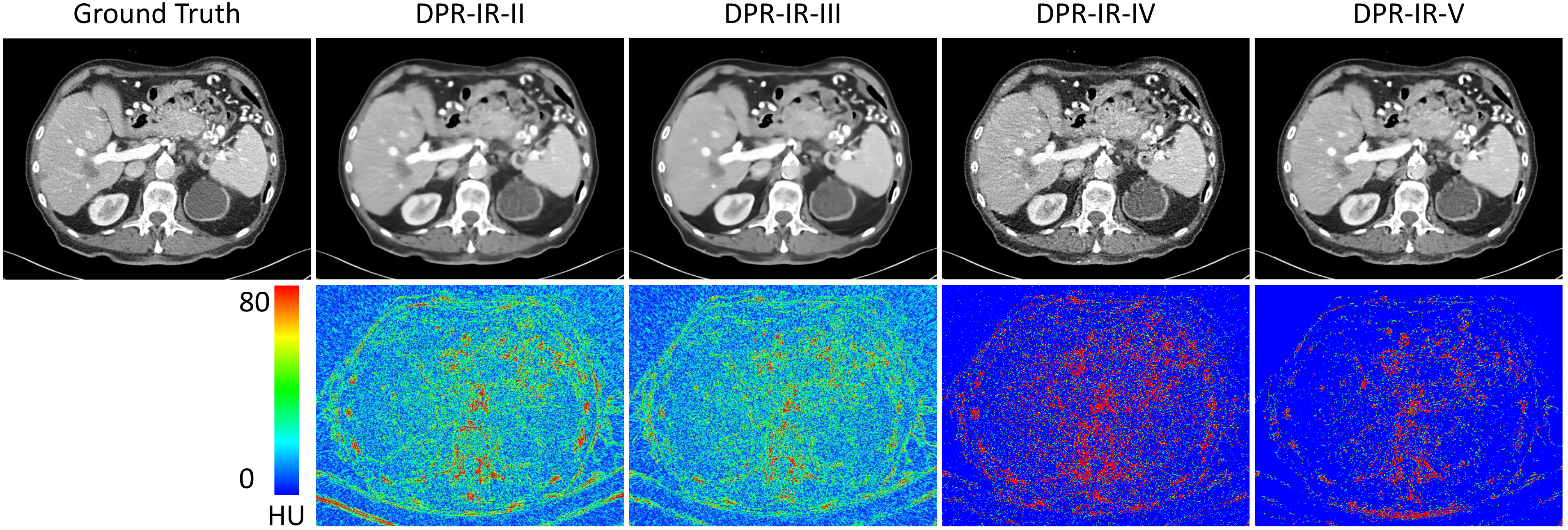}
	\caption{Abdominal reconstructions and their associated absolute difference maps employing 200 sampling steps with 96-view data. The display window is [-160, 240] HU.}
	\label{fig:res_mom_sparseview}
\end{figure*}

Turning to Fig. \ref{fig:res_sparseview}, the streak artifacts have been effectively addressed by DPR-IR. 
Alhough the DPR-IR-V image appears slightly smoother than the DPR-IR-I result, it commendably preserves structures and details. Upon closer inspection of the magnified ROI, DPR-IR-V even retains structures that DPR-IR-I overlooks. Additionally, the DPR-IR-V image exhibits a reduced noise level relative to the DPR-IR-I result, suggesting that the accelerated variant has an enhanced capability to quell noise in the sparse view scenario.

In line with the absolute difference maps, the DPR-IR-V performance is somewhat inferior to the DPR-IR-I counterpart in the low-dose context but slightly excels in the sparse-view setting when judged by quantitative metrics. In general, the two variants produce comparable results, effectively eliminating noise and streaks, while they all retain critical structures. This underscores the efficacy of our CT reconstruction approach, especially our acceleration technique can considerably enhance efficiency without compromising on quality.

\subsection{Conditioning Effects}
This subsection assesses the effects of the conditioning methods on the convergence of the diffusion model-based IR process. Figs. \ref{fig:res_sart_lowdose} and \ref{fig:res_sart_sparseview} present the performance of MCG (GD) and DPR-IR-I. The primary distinction between MCG (GD) and DPR-IR-I lies in the conditioning method.

Regarding the low-dose CT results in Figure 3, the MCG (GD) output exhibits noticeable contour deviations from the ground truth, particularly at the arrow-indicated areas. The difference map reveals considerable discrepancies in the MCG (GD) reconstruction.

A similar pattern can be found from the sparse view results in Figure 4. The MCG (GD) reconstruction suffers from severe internal structural deformations, with significant loss in structural details, especially in arrow-indicated regions. The difference map further emphasizes the substantial errors in the image.

Quantitatively, the metrics on the MCG (GD) image are concerning, with every index underperforming what the competitors offer, even the FBP results. Such anomalies arise from insufficient conditioning since the gradient descent (GD) search is subject to randomness and sub-optimal for convergence. Throughout the sampling process, the stochastic nature of the diffusion model becomes a key factor, leading to substantially distorted results. The structural integrity, intricate details, and sophisticated contours are severely compromised, rendering the results clinically unsatisfactory. In contrast, DPR-IR-I, with the OS-SART method, ensures high convergence efficiency. Every iteration undergoes  data fidelity reinforcement, exerting a positive influence on the final image quality.

These findings underscore the pivotal role of the conditioning method on the convergence of our approach that integrates the diffusion model and the IR algorithm.

\subsection{Momentum Gains}
This subsection delves into the efficacy of our modified Nesterov momentum acceleration, explicitly designed for the diffusion prior. Figs. \ref{fig:res_mom_lowdose} and \ref{fig:res_mom_sparseview} show the outcomes using various techniques. The observations are focused on the clarity and resolution of the reconstruction outcomes.

DPR-IR-II, a momentum-free accelerated version of DPR-IR-I using DDIM, reveals an oversmoothing phenomenon in either the low-dose or sparse-view reconstruction, leading to diminished image resolution. In contrast, employing the conventional momentum acceleration (DPR-IR-III) on $ \bm{x}_t $ does not enhance visual quality; in fact, it exacerbates the oversmoothing effect. With DPR-IR-IV, where momentum is applied to the $ \bm{x}_0 $ estimation in each iteration, there is a marked elevation in image resolution, resulting in superior visual clarity. Yet, in Fig. \ref{fig:res_mom_sparseview}, the DPR-IR-IV outcomes manifest both feature sharpness and heightened noise. These must have come from stark differences between sequential iterations in the diffusion model, causing momentum-induced sharpness and distortion. These issues are notably mitigated with the introduction of the TV constraint on the momentum, as seen in the DPR-IR-V results.

Quantitatively, DPR-IR-III yields the highest scores in terms of the classic image quality metrics, but its 
over-smoothness renders it clinically unsatisfactory. On the other hand, while DPR-IR-V gives slightly reduced scores in these metrics, its perceptual quality is commendable to keep more clinically relevant details and facilite diagnostic tasks.

\subsection{Computational Cost Study}
Table \ref{tab:2} presents the computational time required for CT reconstruction using various methods. Notably, without leveraging acceleration techniques, DPR-IR-I takes approximately 5 minutes per slice. By employing the acceleration method, which reduces the number of sampling steps to 20\% of the original, a near-linear acceleration is observed. Thus, the Nesterov momentum acceleration approach we introduced significantly enhances computational efficiency without compromising performance.

\begin{table}[htbp]
\caption{Computation Cost for Different Methods}
\label{tab:2}
\centering
\begin{tabular}{lcc}
\toprule
\multirow{2}{*}{Method} & \multicolumn{2}{c}{Computational Cost (s)}                            \\
\cmidrule{2-3}
                        & \multicolumn{1}{l}{10\% Dose Data} & \multicolumn{1}{l}{96-View Data} \\
                        \midrule
MCG (GD)                & 296.7                              & 241.7                            \\

DPR-IR-I                & 308.7                              & 284.3                            \\
DPR-IR-II          & 65.3                               & 60.3                             \\
DPR-IR-III               & 64.3                               & 60.7                             \\
DPR-IR-IV              & 65.7                               & 58.0                               \\
DPR-IR-V               & 65.7                               & 60.7  \\
\bottomrule
\end{tabular}
\end{table}

\section{Conclusion}
\label{sec:discussion}
In this paper, we have presented a diffusion prior regularized iterative reconstruction framework. By amalgamating the strengths of the diffusion model and the MBIR algorithm, our approach offers significant advantages. Unlike conventional MBIR algorithms, our approach is empowered by a contemporary image generative model, eliminating the need for any specialized prior design tailored to specific datasets or tasks. A salient feature is its capability for unsupervised learning, eliminating the dependence on labeled data, thereby addressing a prevailing challenge in lacking pairs of clinical images. Our findings have demonstrated the accuracy, robustness and efficiency of our approach, notably in reconstructing images from either low-dose or sparse view data. A pivotal contribution is our innovative momentum acceleration technique tailored for diffusion prior, reducing sampling steps from 1,000 just 200 without compromising image quality.

Our empirical assessment results have suggested that the fusion of the diffusion model and the MBIR algorithm holds a significant promise in the clinical imaging arena. The diffusion model, renowned for its superior image generation prowess, synthesizes images with unparalleled visual realism and clarity. As a result, it delves deeply to systematically glean the prior knowledge of the data, synergistically enhancing the MBIR capability. Reflecting on the triumphs of the MBIR approach over the past two decades, the prospect of this integrated approach should be bright and bring out more advanced imaging solutions.

\bibliographystyle{IEEEtran} 
\bibliography{reference.bib}

\end{document}